\documentclass[twocolumn,amsmath,amssymb,aps,preprintnumbers,showpacs]{revtex4}

\usepackage{ulem}
\usepackage{graphics}% Include figure files
\usepackage{dcolumn}% Align table columns on decimal point
\usepackage{bm}% bold math
\usepackage{amsmath}
\usepackage{amssymb}
\usepackage{hypernat}
\usepackage[ps2pdf,hyperindex=true,final=true,colorlinks=true,pagebackref=false,bookmarks=true,bookmarksopen=true,bookmarksnumbered=true]{hyperref}
\usepackage{epsfig}

\newif\iffig
\newif\ifgraph

\figtrue                     %
\begin{document}

\title{Improving the performance of superconducting microwave resonators in magnetic fields}

\author{D.~Bothner}
\author{T.~Gaber}
\author{M.~Kemmler}
\author{D.~Koelle}
\author{R.~Kleiner}
\affiliation{%
Physikalisches Institut -- Experimentalphysik II and Center for Collective Quantum Phenomena, Universit\"{a}t T\"{u}bingen, Auf der Morgenstelle 14, D-72076 T\"{u}bingen, Germany
}%
\date{\today}

\begin{abstract}
The operation of superconducting coplanar waveguide cavities, as used for circuit quantum electrodynamics and kinetic inductance detectors, in perpendicular magnetic fields normally leads to a reduction of the device performance due to energy dissipating Abrikosov vortices.
We experimentally investigate the vortex induced energy losses in such Nb resonators with different spatial distributions of micropatterned pinning sites (antidots) by transmission spectroscopy measurements at 4.2~K.
In comparison to resonators without antidots we find a significant reduction of vortex induced losses and thus increased quality factors over a broad range of frequencies and applied powers in moderate fields.
\end{abstract}

% PACS, the Physics and Astronomy Classification Scheme.

% 74.25.Qt Vortex lattices, flux pinning, flux creep
%%%%%%%%%% 74. Superconductivity (for superconducting devices, see 85.25.-j)  %%%%%%%%%%

% 74.60.Ge Flux pinning, flux creep, and flux-line lattice dynamics
%%%%%%%%%% 74.78.-w Superconducting films and low-dimensional structures

% 84.40.Dc Microwave circuits

%%%%%%%%%%%%%%%%% 85.25.-j Superconducting devices

%\keywords{Suggested keywords}%Use showkeys class option if keyword display desired

\maketitle

During the last decade coplanar microwave cavities made of superconducting thin films have attained an increasing importance for various experiments and applications.
In circuit quantum electrodynamics, they form besides superconducting artificial atoms/qubits \cite{Clarke08} the elementary building blocks for the fundamental investigation of light-matter interaction on a chip \cite{Wallraff04, Hofheinz09, Niemczyk10}.
These integrated systems have also shown to be suitable candidates for quantum information processing \cite{DiCarlo09}.
As low energy losses, that is high quality factors, are an essential requirement to these resonators, there are currently many efforts to identify and minimize the different dissipation mechanisms \cite{Wang09, Lindstroem09, Macha10, Barends10}. 
Recently, even more advanced hybrid systems have been proposed \cite{Rabl06, Imamoglu09, Verdu09, Henschel10, Bushev10}, coupling real atoms, molecules or electrons to superconducting microwave cavities or combining artificial and real atoms.
Here, microscopic particles need to be trapped and manipulated in the vicinity of the resonator, thereby often requiring external magnetic fields.
Operating superconducting resonators in magnetic fields can lead to considerable energy dissipation due to Abrikosov vortex motion \cite{Song09} and therefore lower quality factors.
Recently there have been some first approaches to overcome this problem under special experimental conditions.
If experimentally feasible, the magnetic field can be applied parallel to the thin film.
This enabled the coupling of spin ensembles in diamond and ruby to superconducting cavities in applied magnetic fields of $\sim100$\,mT \cite{Schuster10}.
For the case of residual ambient fields it has been shown, that energy losses due to a small number of vortices, caught while cooling through the superconducting transition temperature, can be reduced by trapping the vortices within a slot patterned into the resonator \cite{Song09a}.
However, for magnetic fields with a considerable component (mTesla up to Tesla) perpendicular to the superconducting chip, these approaches will not be sufficient.
In this letter we report on the experimental investigation of a method, which leads to a significant reduction of microwave losses in superconducting Nb resonators in perpendicular magnetic fields, as required e.g. for the trapping of ultracold atom clouds on a chip \cite{Fortagh05} or electrons in planar Penning traps \cite{Bushev08}.
We use strategically placed micropatterned holes (antidots) in the superconducting film to provide well-known and highly controllable pinning sites for Abrikosov vortices \cite{Fiory78, Moshchalkov98, Woerdenweber04}.
The presented results are also transferable to other superconducting microwave thin film devices, e.g. kinetic inductance detectors, mixers and filters, when operated in external magnetic fields.

\begin{figure}[h]
\centering {\scalebox{0.41}[0.41]{\includegraphics{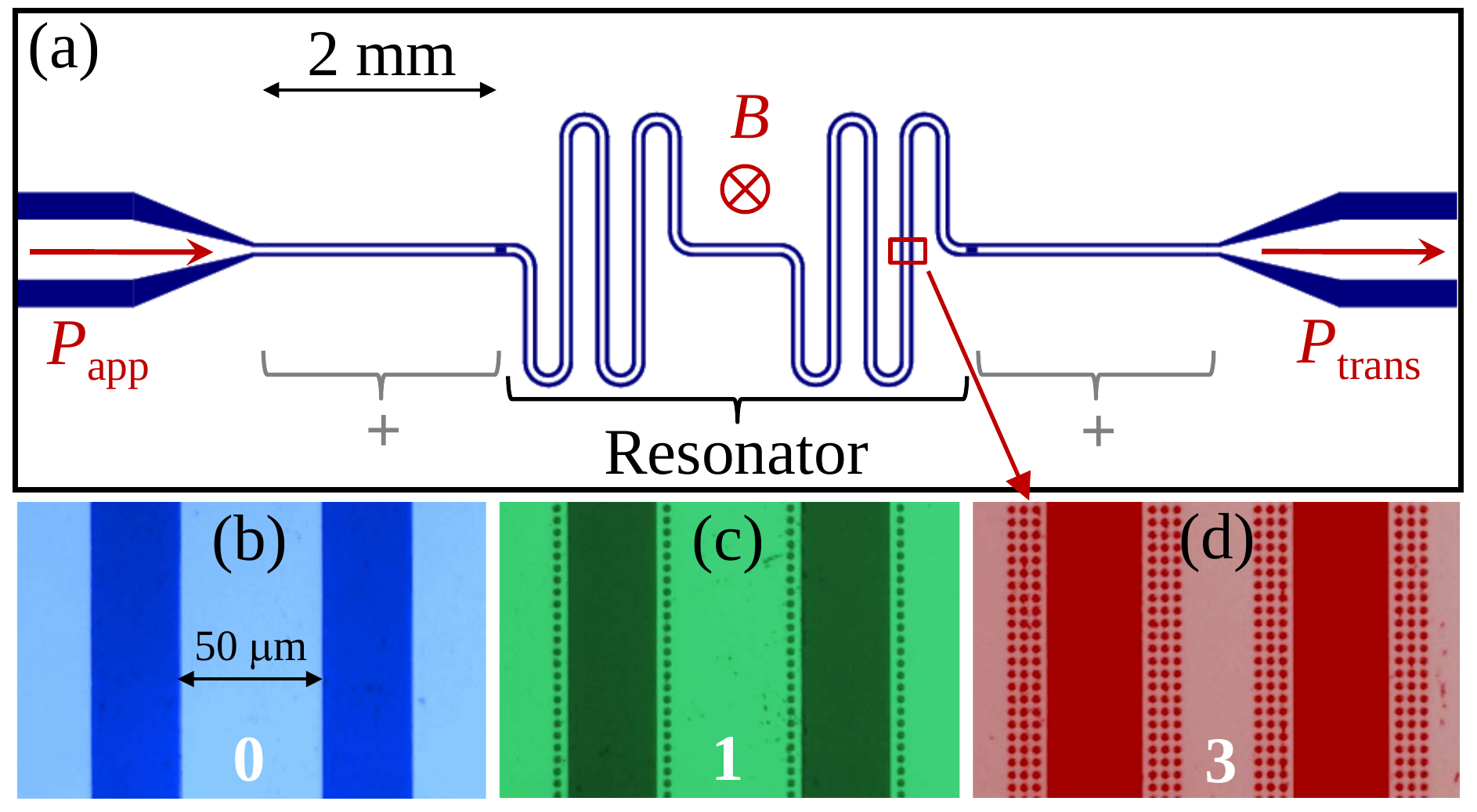}}}
\caption{(Color online) Layout of a $12\times4$\,mm$^2$ chip with a capacitively coupled 3.3 GHz transmission line resonator (a), and optical images of resonators with {\bf0} (b), {\bf1} (c), and {\bf3} (d) rows of antidots.}
\label{fig:Graph1}
\end{figure}
We fabricated half wavelength transmission line resonators with a resonance frequency $f_{\rm{res}}\approx 3.3$\,GHz, capacitively and symmetrically coupled to feed lines via $70$\,$\mu$m wide gaps at both ends.
Figure \ref{fig:Graph1} (a) shows a sketch of the resonator layout.
As the oscillating supercurrents are expected to mainly flow at the edges of the resonator, Abrikosov vortices located there will experience a larger driving force than vortices far away from the edges and therefore give a larger contribution to the losses.
Furthermore, if the magnetic field $B$ is applied with the resonators in the superconducting state, as always in this work, the vortex density will be higher at the edges, where the flux enters \cite{Brandt93}.
Hence, we placed the antidots at the edges of the center conductor and the ground planes in zero (reference sample), one and three rows, denoted as (resonator type) $\bf{0}$, $\bf{1}$ and $\bf{3}$, cf. Fig. \ref{fig:Graph1} (b), (c) and (d).
The antidots have a diameter $d=2\,\rm{\mu m}$ and an antidot-antidot distance $a=4\,\rm{\mu m}$ as design parameters.
As vortices in the feed lines will also contribute to the overall losses and lower the (loaded) quality factor $Q(B)$, we implemented two different designs for resonators $\bf{1}$ and $\bf{3}$: one with antidots only on the resonator ($\bf{1}$, $\bf{3}$) and one with additional antidots on the feed lines ($\bf{1+}$, $\bf{3+}$), cf. Fig. \ref{fig:Graph1} (a).
The antidots on the feed lines have the same configuration as on the respective resonator.
All structures were fabricated on a single $330\,\mu$m thick 2 inch sapphire wafer (r-cut) by optical lithography, dc magnetron sputtering of a 300~nm thick Nb film and lift-off patterning.
The Nb has a critical temperature of $T_c\approx 9$\,K and a residual resistance ratio of $R_{300\rm{\,K}}/R_{10\rm{\,K}}\approx 3.6$. 
The whole wafer was cut into single chips of $12\times4$\,mm$^2$.
Each chip containing one resonator was mounted in a brass box and contacted with Indium to SMA stripline connectors.
After zero-field cooling to $T=4.2$\,K, we measured the frequency dependent transmitted power $P_{\rm{trans}}$ with a spectrum analyzer for different values of applied magnetic field $|B|\leq 4$\,mT, which is perpendicular to the resonator chip.
Due to flux focusing we estimate the flux density seen by the resonator to be one order of magnitude larger than the applied field.
Consequently, the results presented below might be applicable to much higher applied fields, if the ground plane area is properly reduced.
\begin{figure}
\centering \rotatebox{0}{\scalebox{0.42}[0.42]{\includegraphics{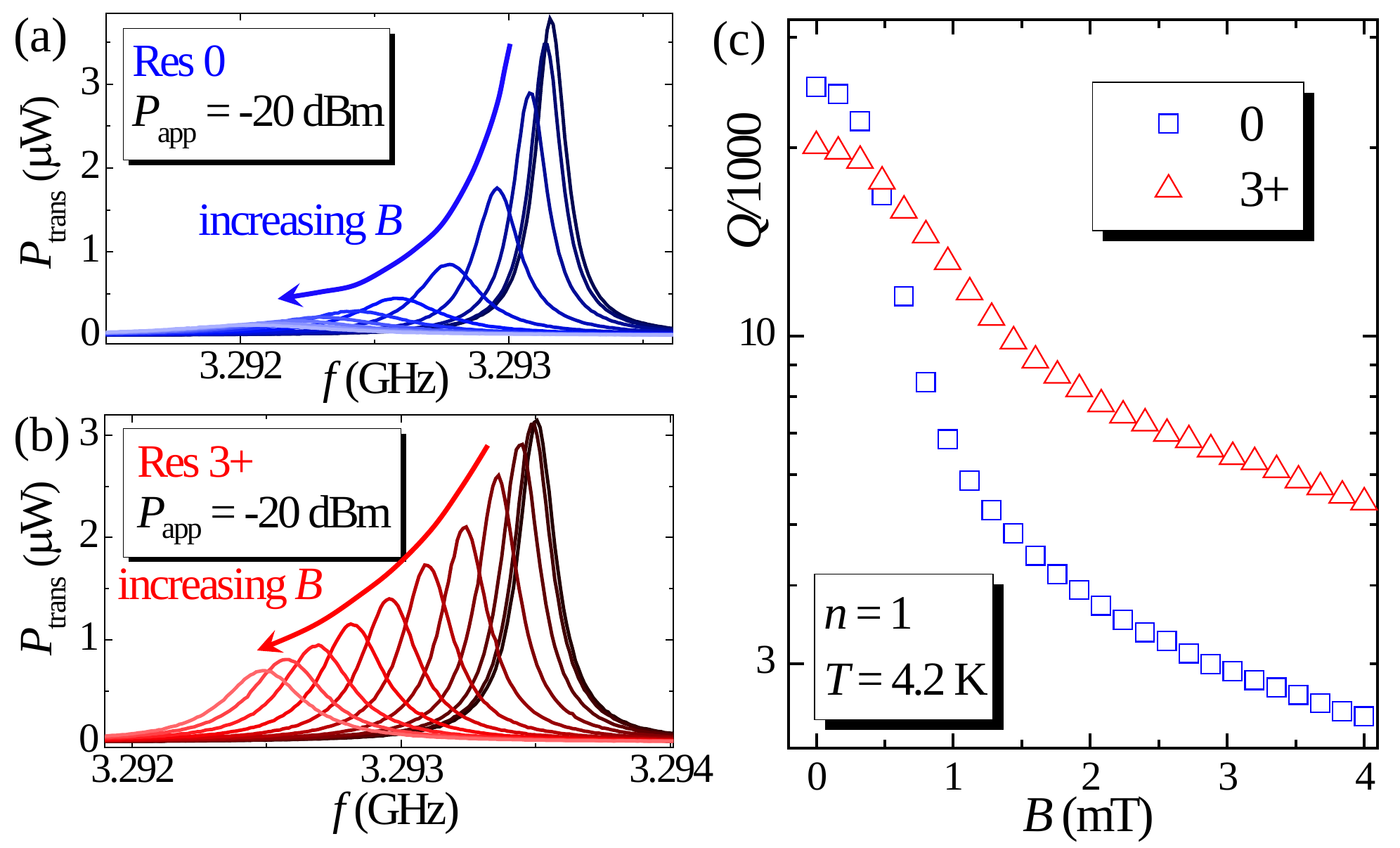}}}
\caption{(Color online) Transmitted power $P_{\rm{trans}}$ vs. frequency $f$ of (a) resonator $\bf{0}$ and (b) resonator $\bf{3+}$ for magnetic fields $0\leq B\leq 1.6$\,mT (in $0.16$\,mT steps); (c) shows corresponding quality factors $Q(B)$ up to $B=4$\,mT.}
\label{fig:Graph2}
\end{figure}
Figure \ref{fig:Graph2} shows resonance curves $P_{\rm{trans}}(f)$ for the fundamental mode $n=1$ of resonator $\bf{0}$ (a) and of resonator $\bf{3+}$ (b) for different values of $B$ and fixed applied power $P_{\rm{app}}$.
For both resonators $f_{\rm{res}}$ shifts to lower frequencies with increasing magnetic field and the resonance peak becomes smaller and broader.
The decrease of transmitted power and broadening of the resonance peak though is much smaller for resonator $\bf{3+}$, indicating reduced field dependent energy losses compared to resonator $\bf{0}$.
For a quantitative analysis we determined resonance frequency $f_{\rm{res}}(B)$ and FWHM $\Delta f(B)$ by fitting the curves with a Lorentzian.
We subsequently calculated the quality factor $Q(B)=f_{\rm{res}}(B)/\Delta f(B)$, shown in Figure \ref{fig:Graph2} (c) for the resonators $\bf{0}$ (blue squares) and $\bf{3+}$ (red triangles).
At $B=0$ the resonator without antidots has the higher $Q(0)$.
However, for $B\geq0.5$\,mT the quality factor of resonator $\bf{3+}$ significantly exceeds $Q(B)$ of resonator $\bf{0}$ up to a factor of $\sim 2.5$ at $B=4$\,mT.
We attribute this enhancement to an effective trapping and pinning of vortices by the antidots.
\begin{figure}[h]
\centering \rotatebox{0}{\scalebox{0.45}[0.45]{\includegraphics{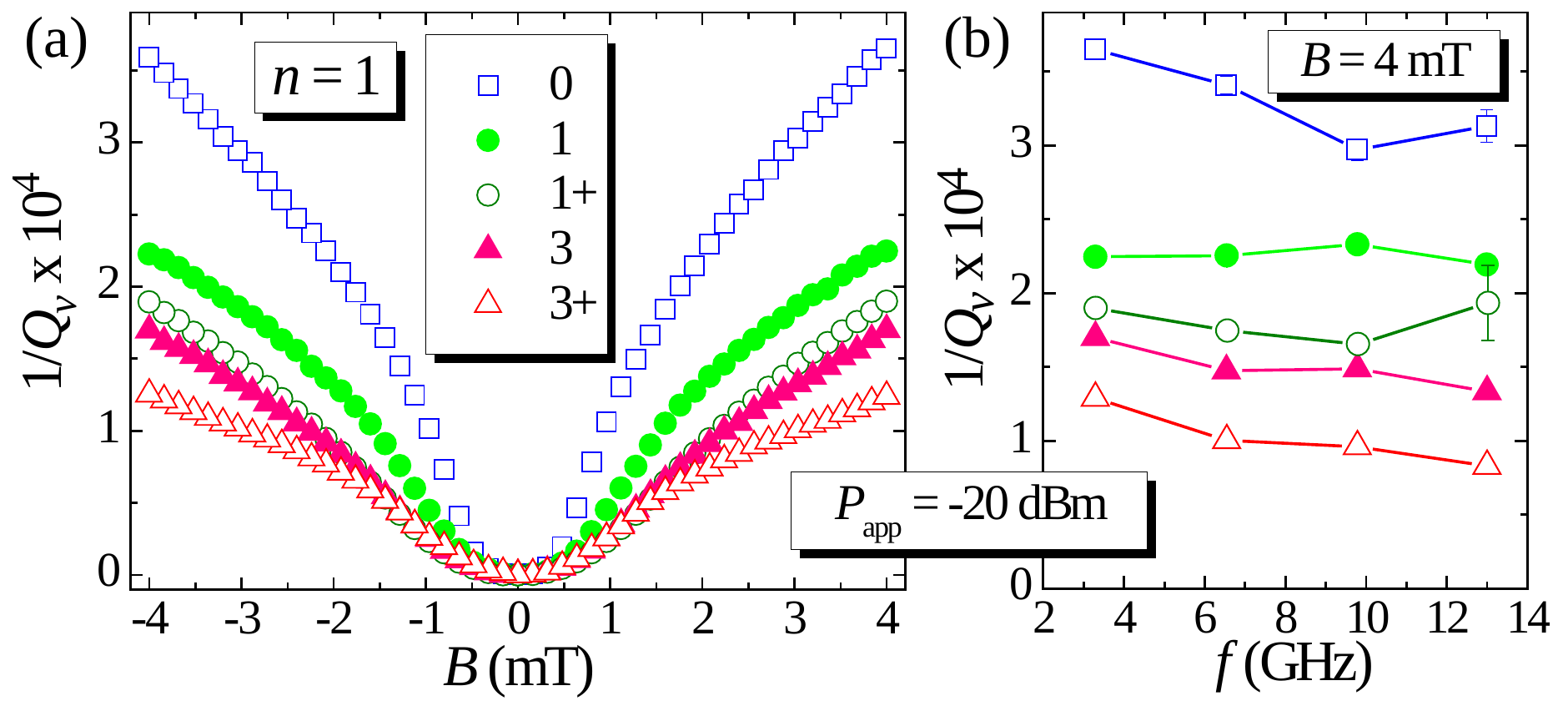}}}
\caption{(Color online) Vortex associated energy loss $1/Q_v$ of five resonators with different antidot distributions (a) vs. applied field $B$ for fundamental mode ($n=1$, $f_{\rm{res}}\approx3.3$\,GHz) and (b) vs. frequency $f$ ($n=1,2,3,4$) at $B=4$\,mT.}
\label{fig:Graph3}
\end{figure}\newline
In general many different field-dependent and field-independent mechanisms contribute to the total energy loss $1/Q(B)$. 
To be able to quantitatively compare the different antidot structures with respect to their ability to suppress vortex induced energy dissipation, we eliminated field independent factors via $1/Q_v(B)=1/Q(B)-1/Q(0)$, cf. Refs. \onlinecite{Song09, Song09a}.
Figure \ref{fig:Graph3} (a) shows the vortex associated energy losses $1/Q_v(B)$ for $n=1$ of five different resonators.
The losses of all resonators with antidots are significantly smaller than the losses of resonator $\bf{0}$, although the magnitude of the reduction varies for the different antidot arrangements.
The reduction of losses increases with increasing number of antidots; three rows of antidots yield about twice the effect of one row of antidots.
The scaling of the loss reduction with the number of antidot rows probably mirrors the nonuniform current and vortex distribution across the resonator: vortices near the edges contribute more to the losses than those further away.
The resonators $\bf{1+}$ and $\bf{3+}$ (open circles and triangles) show even lower losses compared to their counterparts $\bf{1}$ and $\bf{3}$ without pinning sites on the feed lines (full circles and triangles).
This suggests that the ac vortex resistivity in the feed lines has a considerable impact on the overall losses and can be reduced by suitable pinning sites.
We note here, that the quality factor of the five resonators in zero magnetic field varied between $Q(0)\approx15000$ ($\bf{1}$) and $Q(0)\approx43000$ ($\bf{1+}$), but we neither found a correlation between zero field quality factor and antidot configurations nor between $Q(0)$ and $1/Q_v(B)$.
This leads to the conclusion, that the observed reduction of $1/Q_v(B)$ can purely be attributed to effective vortex pinning by the antidots.
We also determined the losses $1/Q_v(B)$ for the first three higher order harmonics $n=2, 3, 4$.
Figure \ref{fig:Graph3} (b) shows $1/Q_v(f)$ for $n=1, 2, 3, 4$ of the five resonators at $B=4$\,mT.
We found that $1/Q_v(f)$ is nearly constant for all five resonators, with a small and surprising tendency to decrease with frequency in contradiction to standard models for the microwave response of vortices \cite{Gittleman68, Brandt91, Coffey91}, which predict a monotonous increase of the resistance with frequency.
A detailed analysis of the data in terms of these models will be given elsewhere.
Despite this somewhat surprising behaviour it is important to note, that the reduction of vortex associated losses by the introduction of antidots is quantitatively and qualitatively stable over a broad frequency range at least from $3.3$\,GHz to $13$\,GHz.
\begin{figure}[h]
\centering \rotatebox{0}{\scalebox{0.47}[0.47]{\includegraphics{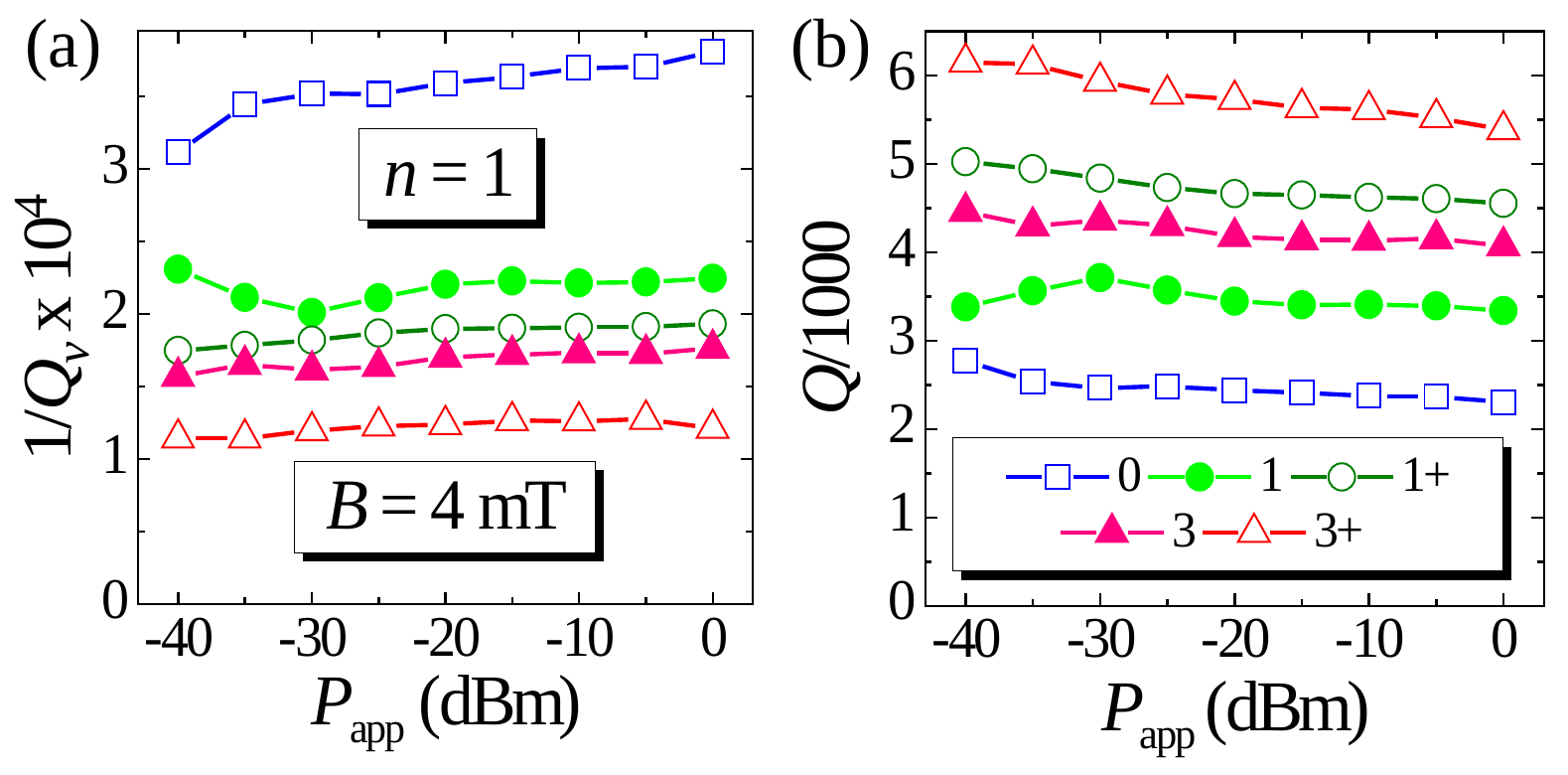}}}
\caption{(Color online) Experimentally determined energy loss $1/Q_v$ (a) and quality factor $Q$ (b) vs. applied microwave power $P_{\rm{app}}$ at $B=4$\,mT and $T=4.2$\,K of five resonators with different antidot distributions.}
\label{fig:Graph4}
\end{figure}
We finally determined the power dependence of the quality factors $Q(B)$ and energy losses $1/Q_v(B)$ for applied powers from $-40$\,dBm to $0$\,dBm.
For $P_{\rm{app}} > 0$\,dBm the resonance peak is cut off, indicating the ac supercurrents to be overcritical near the maximum of the resonance.
For smaller values of $P_{\rm{app}}$ the resonance curves showed no indication of nonlinearities.
Figure \ref{fig:Graph4} depicts (a) $1/Q_v(P_{\rm{app}})$ and (b) $Q(P_{\rm{app}})$ for the five different resonators at $B=4$\,mT.
Both quantities are almost independent of $P_{\rm{app}}$ with a slight tendency of $1/Q_v$ to increase and $Q$ to decrease with increasing applied power.
Note that the absolute quality factor $Q(4$\,mT) is dominated by the vortex induced losses, as it mirrors the behaviour of $1/Q_v(4\,\rm{mT})$ almost independently of $Q(0)$.
$Q(0)$ is nevertheless somewhat perceivable at $B=4$\,mT.
Resonator $\bf{1+}$ for example shows a higher $Q$ than resonator $\bf{3}$ despite the higher vortex associated losses of resonator $\bf{1+}$.
Essentially, the ratios of the vortex losses and the quality factors of the five resonators and therefore the effectiveness of the antidots are almost constant for the investigated power range.
In conclusion, we demonstrated experimentally, that energy losses in superconducting microwave resonators due to the presence of Abrikosov vortices can be significantly reduced by the introduction of antidots.
Accordingly, the quality factor $Q(B)$ at finite applied fields in the mT range can be considerably increased with this method.
We have demonstrated this result to hold for a broad frequency range $3.3$\,GHz $\leq f\leq 13$\,GHz and four orders of magnitude of the applied power $-40$\,dBm $\leq P_{\rm{app}}\leq 0$\,dBm at a temperature of $T=4.2$\,K.
Strategies for transfering these results to higher magnetic fields might include the proper reduction of flux focusing ground planes as well as the implementation of pinning arrays with typical lenghth scales well in the submicron range.
As for many experiments in circuit quantum electrodynamics the resonators are operated in the Millikelvin and single photon regime, the effects presented here have to be investigated under these experimental conditions in further studies.
This work has been supported by the Deutsche Forschungsgemeinschaft (DFG) via the SFB/TRR 21.
D. Bothner acknowledges support by the Evangelisches Studienwerk Villigst e.V..
M. Kemmler acknowledges support by the Carl-Zeiss Stiftung.
We thank Stefan W\"unsch from the Karlsruhe Insitute of Technology (KIT) for sharing his expertise in fabrication and characterization of superconducting resonators, and we thank Roger W\"ordenweber for inspiring discussions.
\bibliography{Bothner_arXiv}
% Produces the bibliography via BibTeX.

\end{document}